\documentclass[aps,prl,
 amsmath,amssymb,
 twocolumn, groupedaddress,
 showpacs]{revtex4-1}
\usepackage{bm}
\usepackage{graphicx}
\usepackage{gensymb}
\usepackage{amsmath}
\usepackage{amssymb}
\usepackage{color}
\usepackage{ulem}

\begin{document}

\title{Unusual temperature evolution of band structure of Bi(111) studied by angle-resolved photoemission spectroscopy and density functional theory}
\author{Takafumi Sato,$^{1,2,3}$ Keiko Yamada,$^2$ Takao Kosaka,$^4$ Seigo Souma,$^{1,3}$ Kunihiko Yamauchi,$^4$ Katsuaki Sugawara,$^{1,2,3}$ Tamio Oguchi,$^4$ and Takashi Takahashi$^{1,2,3}$}

\affiliation{$^1$WPI Research Center, Advanced Institute for Materials Research, Tohoku University, Sendai 980-8577, Japan\\
$^2$Department of Physics, Tohoku University, Sendai 980-8578, Japan\\
$^3$Center for Spintronics Research Network, Tohoku University, Sendai 980-8577, Japan\\
$^4$Institute of Scientific and Industrial Research, Osaka University, Ibaraki, Osaka 567-0047, Japan
 } 

\date{\today}

\begin{abstract}
We have performed angle-resolved photoemission spectroscopy of Bi(111) thin films grown on Si(111), and investigated the evolution of band structure with temperature. We revealed an unexpectedly large temperature variation of the energy dispersion for the Rashba-split surface state and the quantum-well states, as seen in the highly momentum-dependent energy shift as large as 0.1 eV. A comparison of the band dispersion between experiment and first-principles band-structure calculations suggests that the interlayer spacing at the topmost Bi bilayer expands upon temperature increase. The present study provides a new pathway for investigating the interplay between lattice and electronic states through the temperature dependence of band structure.
\end{abstract}


\maketitle

\section{INTRODUCTION}
The interplay between spin-orbit coupling (SOC) and exotic physical properties is one of central topics in condensed-matter physics, as exemplified by the intensive investigation of topological insulators (TI) \cite{HasanReview, SCZhangReview, AndoReview} and noncentrosymetric/Rashba superconductors \cite{Sigrist2004}, where the SOC-induced spin-split energy band plays a central role in characterizing the spin-helical Dirac fermions and the unconventional superconducting pairing. Group-V semimetal bismuth (Bi) is a key element to investigate the spin-orbit-coupled electronic states and its relationship with physical properties, owing to the fairly strong SOC originating from the heavy atomic mass of Bi. Besides the fundamental interest of bulk Bi for valleytronics \cite{ZhuNP2012, KuchlerNM2014}, the strong SOC of Bi leads to several exotic quantum states, as highlighted by the Rashba spin splitting associated with the broken space-inversion symmetry at the surface \cite{KoroteevPRL2004, HofmannPSS2006}, the quantum spin-Hall-insulator phase in one to a few bilayers (BLs) of Bi(111) \cite{MurakamiPRL2006, LiuPRL2014, YazdaniNP2014}, and non-trivial topological phases in Bi-based compounds such as Bi$_{1-x}$Sb$_x$, Bi$_2$Se$_3$, and Bi$_2$Te$_3$ \cite{FuPRB2007, ZhangNP2009}.

 Angle-resolved photoemission spectroscopy (ARPES) has played a central role in clarifying the electronic states of bulk and thin-film Bi, by observing the Rashba-spin-split surface state (SS) and its momentum-locked spin texture \cite{AstPRL2001, AstPRL2003, KoroteevPRL2004, HofmannPSS2006, HiraharaPRL2006, HiraharaPRB2007, HiraharaNJP2008, TakayamaPRL2011, OhtsuboPRL2012, TakayamaNL2012, YamadaNL2018, ShimamuraACSNano2018, ItoSciAdv2020}. Bulk Bi, known as a typical low-carrier semimetal, crystalizes in the rhombohedral A7 structure and has a BL-terminated structure stacked along the [111] direction \cite{HofmannPSS2006}, as shown in Fig. 1(a). ARPES studies of Bi(111) thin films further clarified the quantum-well states (QWS), whose energy dispersion and spin polarization strongly depend on the film thickness due to the quantum-size effect \cite{HiraharaPRL2006, HiraharaPRB2007, HiraharaNJP2008, KoroteevPRB2008, TakayamaPRL2011, OhtsuboPRL2012, TakayamaNL2012, YamadaNL2018, ShimamuraACSNano2018, ItoSciAdv2020}. As revealed by the intensive ARPES studies, a small indirect overlap in the semimetallic band structure makes the electronic state of Bi(111) very sensitive to structural, electronic, and chemical parameters, such as the film thickness $d$, chemical potential, SOC, and surface condition \cite{HiraharaPRL2006, HiraharaPRB2007, HiraharaNJP2008, KoroteevPRB2008, TakayamaPRL2011, OhtsuboPRL2012, TakayamaNL2012, YamadaNL2018, ShimamuraACSNano2018, ItoSciAdv2020, HsiehNature2008, HsiehScience2009, NishidePRB2010, GuoPRB2011}. While electronic phase transitions (such as metal to semiconductor or TI to ordinary insulator) could be induced by tuning some of above parameters in Bi, the evolution of electronic states as a function of temperature has been scarcely explored, except for a few studies \cite{HofmannPRL2003, HofmannPSS2006} reporting the electron-phonon-coupling-induced lifetime broadening. This is presumably because a strong temperature dependence of electronic states was not expected in this system due to the essentially weak electron correlation of the Bi $sp$ orbital and the absence of temperature-induced phase transition \cite{MonigPRB2005}.
 
 In this article, we investigated the electronic states of Bi(111) thin film by ARPES, and uncovered a marked temperature dependence of the band structure which was overlooked in previous studies. We also performed first-principles band-structure calculations for Bi(111) slabs, and found that the observed temperature evolution of the band structure is ascribed to the temperature-induced variation in the interlayer spacing in the topmost BL in the film.

\section{EXPERIMENT AND CALCULATION}
 We explain how to prepare a Bi thin film. At first, we heated a Si(111) substrate at 1000$^\circ$C to obtain a clean well-ordered 7$\times$7 surface, and deposited Bi atoms onto the Si substrate at room temperature by using a Knudsen cell equipped in the molecular-beam-epitaxy chamber with the base pressure of better than 5 $\times$ 10$^{-10}$ Torr. Subsequently, the film was annealed at 150$^\circ$C to improve the crystallinity. The film thickness ($d$) was controlled by the deposition time with the constant deposition rate (0.09 \AA/s), and estimated by the quartz-oscillator thickness monitor and the energy position of the QWS in ARPES spectrum \cite{HiraharaPRL2006, HiraharaPRB2007, HiraharaNJP2008, TakayamaPRL2011}. The 1$\times$1 surface structure was confirmed by the low-energy electron diffraction (LEED) measurement. ARPES measurements were performed using an MBS-A1 electron energy analyzer equipped with a xenon plasma discharge lamp. We used the Xe-I resonance line (photon energy $h\nu$ = 8.437 eV) and a vacuum-ultraviolet continuous-wave (CW) laser (LEOS solutions) with $h\nu$ = 6 eV to excite photoelectrons. The energy resolution was set to be 4-30 meV. Band-structure calculations were carried out by means of the first-principles density-functional-theory (DFT) approach using the HiLAPW code with the all-electron full-potential linearized augmented-plane-wave method in a scalar-relativity plus SOC manner. In the calculations, a thin-film was simulated by adopting free-standing periodic slabs with centrosymmetric P$\bar{3}m$1 space group with vacuum spacing of 15 \AA.

\section{RESULTS AND DISCUSSION}

\begin{figure}
\begin{center}
\includegraphics[width=3.2in]{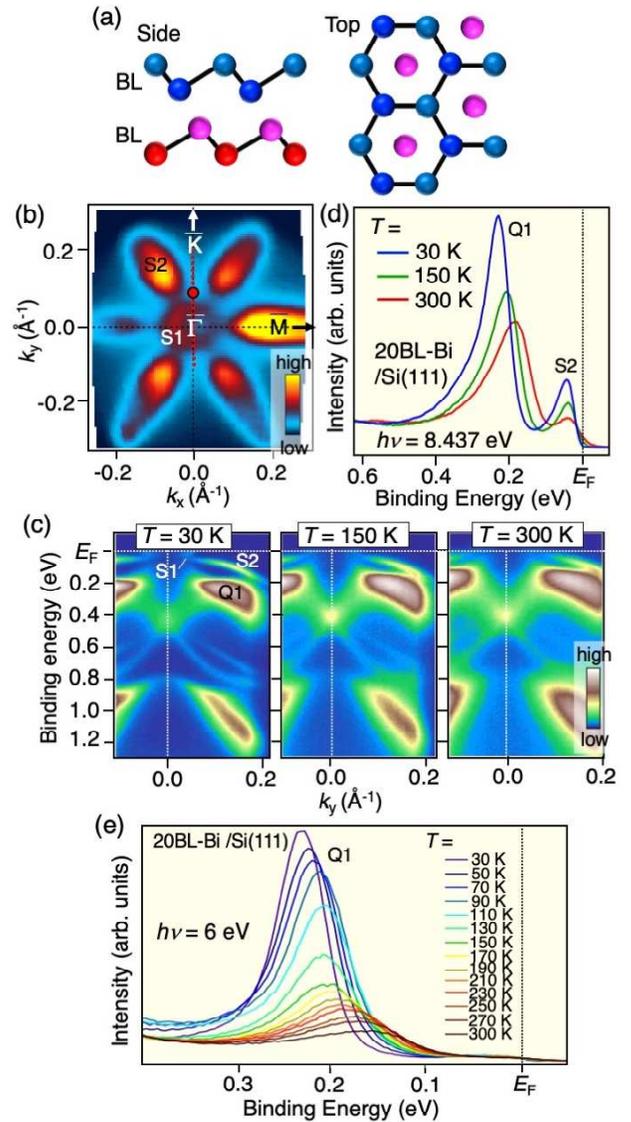}
  \hspace{0.2in}
  \caption{(color online). (a) Top and side views of crystal structure of Bi(111). (b) ARPES-intensity mapping at $E_{\rm F}$ as a function of 2D wave vector ($k_x$ and $k_y$) for 20BL Bi(111) on Si(111). (c) ARPES-intensity plots as a function of $k_y$ and $E_{\rm B}$ in a relatively wide $E_{\rm B}$ region at $T$ = 30, 150, and 300 K, measured along the $\bar{\Gamma}\bar{K}$ cut shown by a red line in (b). (d) EDC at three representative temperatures ($T$ = 30, 150, and 300 K) measured with the Xe-I line ($h\nu$ = 8.437 eV) at the $k$ point indicated by a red dot in (b). (e) Temperature dependence of EDC for 20BL-Bi(111)/Si(111) measured at the $k_{\rm F}$ point of the S2 band along the $\bar{\Gamma}\bar{M}$ cut with a 6-eV CW laser. To determine the $k_{\rm F}$ point, we fitted the EDCs around the peak-top with a Lorentzian and estimated the peak position. By smoothly connecting the peak positions with polynomial function, we have determined the experimental band dispersion and the $k_{\rm F}$ point.}
\end{center}
\end{figure}

\begin{figure*}
\includegraphics[width=6.3in]{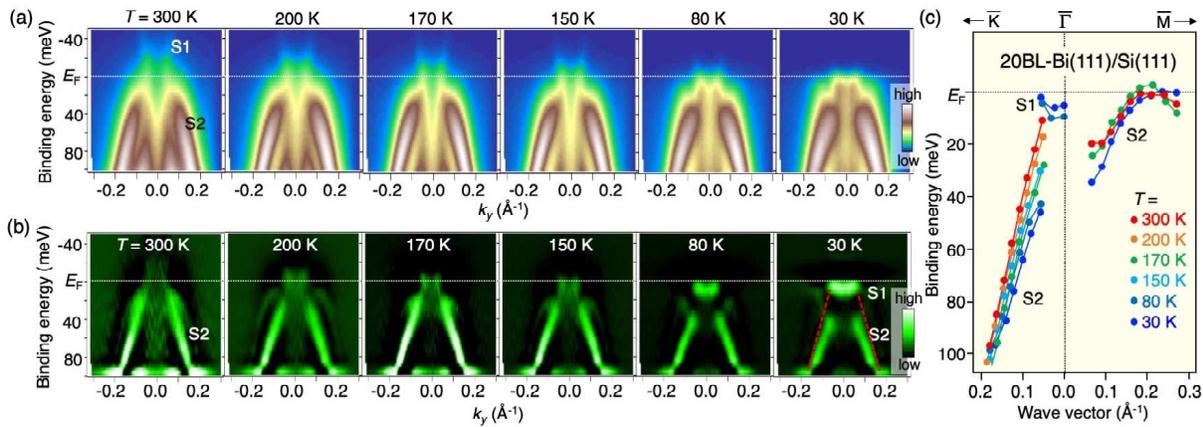}
\hspace{0.2in}
 % \vspace{-0.3in}
\caption{(color online).  (a, b) Temperature dependence of the ARPES intensity and its second derivative intensity of EDCs, respectively, for 20BL-Bi(111)/Si(111) measured around the $\bar{\Gamma}$ point along the $\bar{\Gamma}\bar{K}$ cut with the Xe-I line. (c) Experimental band dispersion along the $\bar{\Gamma}\bar{K}$ and $\bar{\Gamma}\bar{M}$ cuts at various temperatures (solid curves), extracted by tracing the peak position of EDCs at each temperature (squares).}
\end{figure*}

Figure 1(b) shows the ARPES-intensity mapping at $E_{\rm F}$ at $T$ = 30 K for 20BL Bi(111) on Si(111) measured with the Xe-I line. One can immediately recognize two types of Fermi surfaces, a hexagonal electron pocket centered at the $\bar{\Gamma}$ point and six elongated hole pockets surrounding the hexagonal pocket, both of which are assigned to the Rashba SS \cite{KoroteevPRL2004, HofmannPSS2006}. The former pocket originates from a weak feature crossing $E_{\rm F}$ (called S1) in the ARPES intensity along the $\bar{\Gamma}\bar{K}$ cut in Fig. 1(c), while the latter from a sharp topmost holelike band approaching $E_{\rm F}$ toward the $\bar{\Gamma}$ point (called S2). One can see in Fig. 1(c) a few prominent dispersive features at binding energy $E_{\rm B} > $ 0.2 eV at $T$ = 30 K, most of which are the QWSs originating from the quantum confinement of bulk bands. Among these QWSs, we mainly focus on the topmost QWS (called Q1). One can recognize from Fig. 1(c) that the overall spectral feature becomes gradually obscured on increasing temperature from 30 K to 300 K due to the thermal-broadening effect. A careful look at the intensity pattern further reveals that the Q1 band around $k_y \sim$0.1 \AA$^{-1}$  \hspace{0.5mm} slightly moves upward on elevating temperature. This is better visualized in the energy distribution curve (EDC) at such a $k$ point [a red circle in Fig. 1(b)] in Fig 1(d), where a peak from the Q1 band apparently moves toward $E_{\rm F}$ on increasing temperature. This trend is also seen in the near-$E_{\rm F}$ peak originating from the S2 band although the energy shift is much smaller; we will come back to this point later. Such unprecedented temperature dependence cannot be explained in terms of a simple thermal-broadening effect. In fact, we have confirmed by the numerical simulation that an extra gaussian broadening of the EDC at 30 K cannot reproduce the EDC at 300 K. To further examine the intrinsic nature of the band shift, we have performed ARPES with CW laser ($h\nu$ = 6 eV) at the Fermi vector ($k_{\rm F}$) of the S2 band along the $\bar{\Gamma}\bar{M}$ cut with a much finer temperature step, as shown in Fig. 1(e). On increasing temperature from 30 K, one can again recognize a clear shift of the energy position for the Q1 peak toward lower $E_{\rm B}$ by $\sim$0.1 eV, accompanied with a strong broadening of the peak. This suggests that the observed energy shift is not an experimental artifact associated with the experimental condition (such as photon energy and light polarization) in the ARPES measurement.

Now that the temperature-induced band shift is established for the Q1 band, next we examine it for the Rashba SS near $E_{\rm F}$ with a higher experimental precision. Figures 2(a) and 2(b) show the temperature dependence of ARPES intensity and the corresponding second derivative intensity of EDCs in the close vicinity of $E_{\rm F}$, respectively, for 20BL-Bi(111)/Si(111) measured along the $\bar{\Gamma}\bar{K}$ cut around $\bar{\Gamma}$. One can see from the intensity plot at $T$ = 300 K [left-most panel of Fig. 2(a)] that the S2-band-derived holelike dispersion rapidly approaches $E_{\rm F}$ around the $\bar{\Gamma}$ point and the S1-derived electronlike dispersion is well visible above $E_{\rm F}$ due to a finite population of the Fermi-Dirac (FD) function at $T$ = 300 K. This electronlike band becomes less visible at lower temperatures due to the steeper cutoff of FD function [see right panels of Fig. 2(a)]. Besides such a FD-function-related change in the intensity pattern, one can see in the second-derivative intensity plots in Fig. 2(b) a systematic downward shift of the S2 band around the $\bar{\Gamma}$ point on decreasing temperature, as clearly recognized in the intensity plot at $T$ = 30 K showing an apparent deviation of the dispersion from that at $T$ = 300 K (highlighted by a red dashed curve). The marked temperature dependence of the S2 band is better illustrated in the experimental band dispersion extracted from the peak position in EDCs [Fig. 2(c)]. The temperature dependence of the band position along the $\bar{\Gamma}\bar{K}$ cut is not rigid-band-like, and the variation becomes stronger on approaching the $\bar{\Gamma}$ point (maximally $\sim$ 35 meV at $k \sim$ 0.05 \AA$^{-1}$). This trend is also seen in the dispersion along the $\bar{\Gamma}\bar{M}$ cut (note that it is difficult to estimate the band dispersion of the S2 band at the $\bar{\Gamma}$ point because of the sudden intensity drop due to the existence of bulk-band projection). Importantly, the band shift becomes less clear at the $k_{\rm F}$ point of the S1 (S2) band along the $\bar{\Gamma}\bar{K}$ ($\bar{\Gamma}\bar{M}$) cut, suggesting that the location of the $k_{\rm F}$ points is almost unchanged with temperature. This is reasonable in light of the Luttinger theorem because the strong temperature dependence of carrier concentration is not expected and the charge transfer across the interface from Si(111), if it exists, would be negligibly small for the topmost surface probed by ARPES. Thus, the observed temperature-dependent band shift is not associated with the change in the surface charge. The band shift may not be associated with the adsorption of impurities on the surface upon cooling down the sample because (i) the observed strongly momentum-dependent band shift in Fig. 2(c) is incompatible with the rigid-band shift expected from the simple electron-doping scenario associated with the impurity adsorption, and (ii) the peak position in ARPES spectrum at each temperature was observed to be insensitive to the vacuum condition (i.e. impurity-adsorption condition) of the ARPES-measurement chamber. It is also unlikely that the band shift is triggered by a structural phase transition, because such a transition has not been reported for Bi(111) and the observed shift is gradual with temperature, incompatible with the phase-transition scenario. A possibility of surface photo-voltage effect would be also excluded because of the high-metallicity of the surface, unlike the case of Si \cite{DemuthPRL1986}. It may be thus possible to attribute the observed shift to the change in the structural parameter of Bi(111). It is noted here that possible spatial variation of the film thickness does not play a serious role to the observation of band shift and its interpretation, because the band dispersion of the Rashba SS (S1 and S2) is insensitive to the film thickness for $\sim$20BL-thick regime.

We found from the location of the LEED spots as well as the absolute $k$ value of the $\bar{M}$ point in the ARPES result that the in-plane [(111) plane] lattice constant ($a$) exhibits no detectable change with temperature, consistent with the small thermal expansion coefficient of bulk Bi \cite{ErflingAP1939}. Thus, we speculate that the out-of-plane structural parameter (along the [111] axis) of the topmost surface (but not the whole crystal) may change, as suggested by a previous LEED intensity-vs-voltage ($IV$) measurement of Bi(111) \cite{MonigPRB2005}. As shown in Fig. 3(a), there exist two types of out-of-plane structural parameters, i.e. the first interlayer spacing ($d_{\rm 12}$) and the inter-BL distance ($d_{\rm 23}$). We have carried out band-structure calculations for 10BL Bi(111) slab [Fig. 3(a)] to examine the sensitivity of band structure to these two parameters. In the calculations, we fixed the interlayer spacing and the inter-BL distance to the bulk values (1.590 and 2.342 \AA, respectively) for the 2nd-9th BLs \cite{MonigPRB2005, DuNC2016, LiuPRB1995}. We also fixed the in-plane lattice constant $a$ to be 4.533 \AA \hspace{0.5mm} for all the BLs. While varying the $d_{\rm 12}$ and $d_{\rm 23}$ values for the 1st BL (top surface), we have also changed these values for the 10th BL (bottom surface) to avoid undesired symmetry reduction. Figure 3(b) shows the calculated total electron energy plotted as a function of $d_{\rm 12}$ around its local minimum for different $d_{\rm 23}$ values (note that the total energy corresponds to the value for the one-side surface obtained by taking a half value of the energy for the whole supercell). The total energy has a nearly parabolic dependence against $d_{\rm 12}$ and shows a minimum at ($d_{\rm 12}, d_{\rm 23}$) = (1.61 \AA, 2.50 \AA). This minimum is recognized from the plot of total energy against $d_{\rm 23}$ for fixed $d_{\rm 12}$ in Fig. 3(c), whereas the shape of parabola appears to be slightly asymmetric compared to that in Fig. 3(b). It is noted that the optimized values of $d_{\rm 12}$ and $d_{\rm 23}$ at which the total energy takes the minimum are different from the bulk values by 1.2\% and 6.7\%, respectively. This may be caused by a lattice relaxation at the surface due to the abrupt termination of crystal structure.

\begin{figure}
 \includegraphics[width=3.4in]{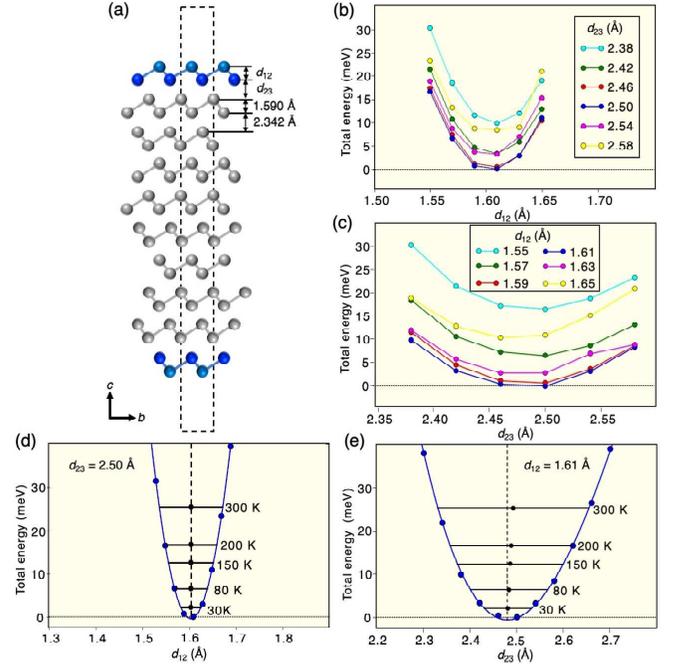}
 \hspace{0.2in}
  \vspace{-0.2in}
\caption{(color online). (a) Free-standing 10BL-Bi(111) slab with a vacuum layer of 15 \AA\hspace{0.5mm} adopted in the band-structure calculations. Dashed rectangle represents the unit cell of an actual slab. Definition of $d_{\rm 12}$ (first interlayer spacing) and $d_{\rm 23}$ (inter-BL distance) is also indicated. Outermost (top and bottom) Bi BLs are shown by colored (light blue and dark blue) circles. The first and second Bi monolayer (both are within single BL) from the surface are indicated by light blue and dirk blue circles, respectively. In the calculations, interlayer spacing within each BL and inter-BL distance are fixed to 1.590 and 2.342 \AA, respectively, for the 2nd-9th BLs (indicated by gray circles). In-plane lattice constant is fixed to $a$ = $b$ = 4.533 \AA. Note that the $a$ value ($\sim$4.5 \AA) was found to be temperature independent in the experiment. (b) Calculated total energy as a function of $d_{\rm 12}$ around the local minimum for various $d_{\rm 23}$ values (2.38-2.58 \AA). (c) Calculated total energy as a function of $d_{\rm 23}$ for representative $d_{\rm 12}$ values (1.55-1.65 \AA). Lowest energy at ($d_{\rm 12}, d_{\rm 23}$) = (1.61 \AA, 2.45 \AA) was set as an origin of the total energy in (b) and (c). (d) Total energy as a function of $d_{\rm 12}$ at $d_{\rm 23}$ = 2.50 \AA (dots) and the result of numerical fitting with a cubic function. Horizontal lines are characteristic energy ($k_{\rm B}T$/2) at representative temperatures ($T$ = 300, 200, 150, 80, and 30 K). Black dot represents an oscillation center at each temperature based on a simple vibration model. (e) Same as (d) but plotted as a function of $d_{\rm 23}$ at $d_{\rm 12}$ = 1.61 \AA. A vertical dashed line in (d) and (e) indicates the oscillation center at $T$ = 30 K. The oscillation center shifts toward larger $d_{\rm 23}$ at higher temperature.}
\end{figure}

To examine a possible temperature-induced change of structural parameter in terms of the anharmonicity of total energy, we numerically fit the total energy vs $d_{\rm 12}$ curve at $d_{\rm 23}$ = 2.50 \AA \hspace{0.5mm} with the cubic function as shown by a solid curve in Fig. 3(d). Then we consider a putative one-dimensional translational movement of atoms along the [111] direction with a kinetic energy term $k_{\rm B}T$/2 to account for a finite temperature effect (a simple oscillator model). The oscillation center at $T$ = 30 and 300 K estimated by taking the midpoint of two intersections between the equi-energy line (horizontal line) and the total-energy curve is $d_{\rm 12}$ = 1.6047 and 1.6052 \AA, respectively (marked by black dots on the vertical dashed line), corresponding to a change of 0.03\% in $d_{\rm 12}$. This value is obviously too small to cause a meaningful change in the band structure. Despite such a harmonic behavior of the total energy against $d_{\rm 12}$, that against $d_{\rm 23}$ was found to exhibit a strong anharmonicity. As shown in Fig. 3(e), the oscillation center gradually moves toward larger $d_{\rm 23}$ on increasing temperature, as visible from a systematic deviation of black dot from a vertical dashed line (corresponding to the oscillation center at $T$ = 30 K) at higher temperatures. Indeed, the estimated oscillation center varies by 0.4\% from 30 K (2.4804 \AA) to 300 K (2.492 \AA), more than one order of magnitude larger than the case for $d_{\rm 12}$. However, we will show later from the calculated band dispersion in Fig. 4(c) that this amount of $d_{\rm 23}$ variation is still insufficient to account for the observed large temperature variation of the band dispersion. 
%%This suggests that some unknown factors which are not explained in terms of the simple anharmonicity of lattice vibration may be responsible for the temperature-dependent lattice relaxation at the surface.

Now we discuss the relationship between the calculated band structure and surface structural parameters. Figure 4(a) shows the calculated band structure along two high-symmetry cuts, $\bar{\Gamma}\bar{K}$ and $\bar{\Gamma}\bar{M}$ for 10BL Bi(111) with the optimized $d_{\rm 12}$ and $d_{\rm 23}$ values for $T$ = 0 K [($d_{\rm 12}$, $d_{\rm 23}$) = (1.61 \AA, 2.50 \AA)]. One can see several bands originating from the quantum size effect. Among these bands, a band indicated by an arrow is assigned to the S2 band because its overall dispersive feature, e.g. the $E_{\rm F}$-crossing along the $\bar{\Gamma}\bar{M}$ cut associated with a small elongated hole pocket and the convex shape of band dispersion along the $\bar{\Gamma}\bar{K}$ cut, are similar to those in the experiment shown in Fig. 2(c). The experimental S1 band is not well reproduced in the calculation probably due to its proximity to the projection of bulk band forming a small hole pocket at the T point of bulk BZ (i.e. at the $\bar{\Gamma}$ point of surface BZ). We have examined the sensitivity of calculated S2-band dispersion in the vicinity of $E_{\rm F}$ to the variation of $d_{\rm 12}$ and $d_{\rm 23}$, and show the results in Figs. 4(b) and 4(c). When $d_{\rm 23}$ is fixed to an optimized value of 2.50 \AA \hspace{0.5mm} and $d_{\rm 12}$ is decreased from 1.65 to 1.59 \AA, the S2 band displays an overall downward shift. This shift is momentum dependent; it is the largest around the $\bar{\Gamma}$ point ($\sim$30 meV) and becomes gradually small on moving toward $\bar{K}$. This trend is similar to the temperature dependence of experimental band structure in Fig. 2(c). On the other hand, when $d_{\rm 12}$ is fixed to an optimized value of 1.61 \AA  \hspace{0.5mm} and $d_{\rm 23}$ is decreased from 2.58 to 2.46 \AA \hspace{0.5mm} [note that this range is much wider than that discussed in Fig. 3(e)], the calculated band structure shows a much weaker change as shown in Fig. 4(c). These results suggest that the energy dispersion is more sensitive to the change in $d_{\rm 12}$ than $d_{\rm 23}$, as supported by the previous DFT calculation \cite
{DuNC2016}. This would be reasonable when we take into account the interlayer bonding within each BL (covalent type) which is much stronger than the inter-BL one (van-der-Waals type), so that even a small change in the interlayer bond length ($d_{\rm 12}$) can effectively alter the energy position of bands. It is thus inferred from the consideration of both the experimental temperature-dependent band shift and the $d_{\rm 12}$-sensitive variation of calculated S2-band dispersion that the interlayer spacing ($d_{\rm 12}$) in the topmost BL expands on increasing temperature, as schematically shown in Fig. 4(d). It is noted that this explanation is speculative and needs be checked by the structural measurements in future, e.g., with transmission electron microscopy (TEM) or selected-area diffraction. It is also emphasized that the simple argument on the temperature dependence of $d_{\rm 12}$ based on the anharmonicity of the total energy as a function of $d_{\rm 12}$ [Fig. 3(d)] is insufficient to explain the much larger variation of $d_{\rm 12}$ ($\sim$3\%; 1.65/1.59 \AA) inferred from the comparison of ARPES data [Fig. 2(c)] with the calculated $d_{\rm 12}$-dependent band structure [Fig. 4(b)]. This suggests that we should take into account an additional factor beyond the anharmonicity of total energy to understand the large variation of $d_{\rm 12}$, such as the coupling of electrons to the phonons associated with the in-plane ionic displacement in the honeycomb lattice within the Bi BL.

\begin{figure}
 \includegraphics[width=3.4in]{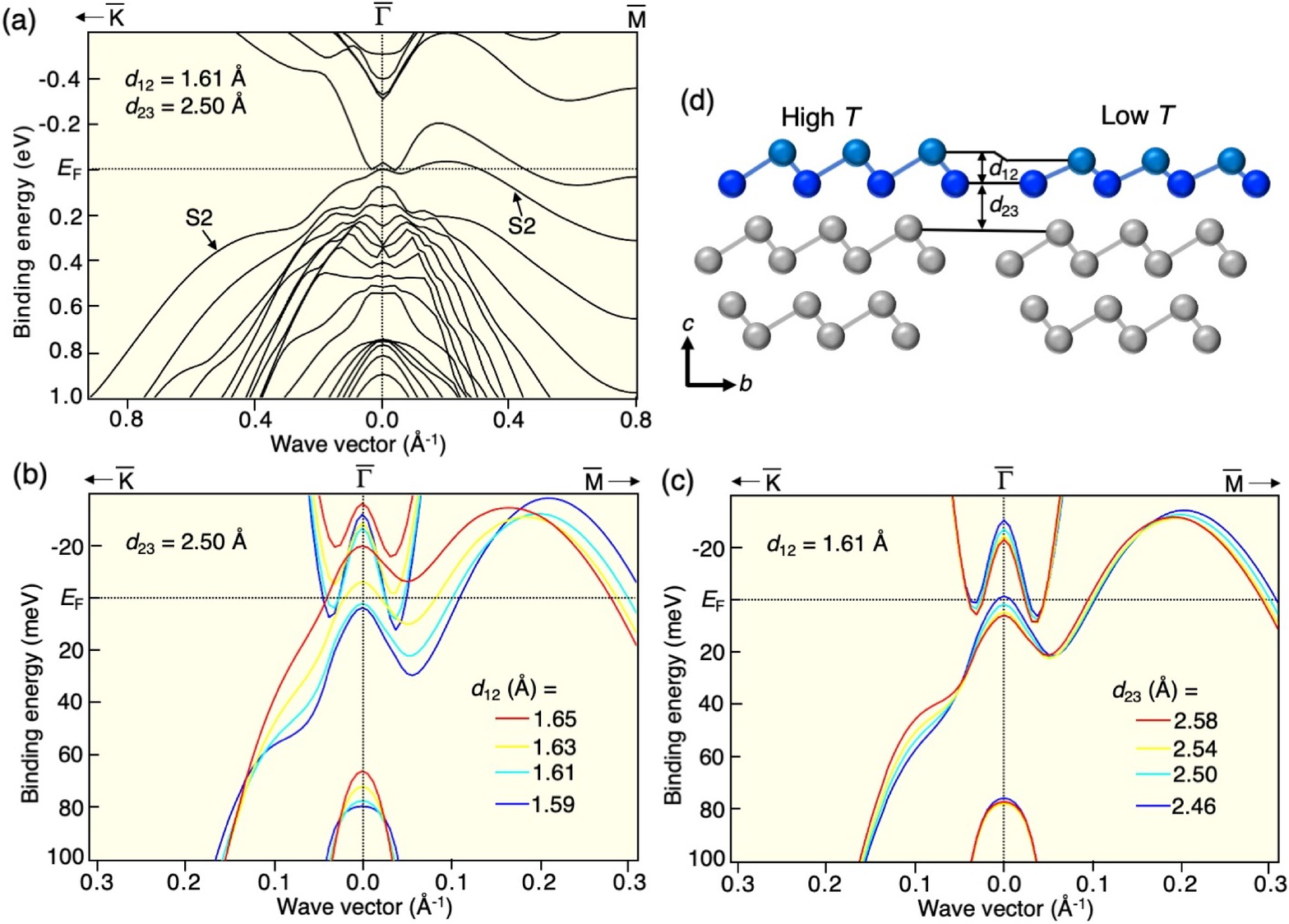}
 \hspace{0.2in}
  \vspace{-0.1in}
\caption{(color online). (a) Calculated band structure in a wide $E_{\rm B}$ region for 10BL Bi(111) slab with ($d_{\rm 12}, d_{\rm 23}$) = (1.61 \AA, 2.50 \AA) (optimized values for $T$ = 0 K). (b) Calculated near-$E_{\rm F}$ band structure for four representative $d_{\rm 12}$ values at fixed $d_{\rm 23}$ ( = 2.50 \AA). (c) Same as (b) on varying $d_{\rm 23}$ at fixed $d_{\rm 12}$ (= 1.61 \AA). The results in (b) and (c) highlight that the calculated band structure is sensitive (insensitive) to the variation in $d_{\rm 12}$ ($d_{\rm 23}$). (d) Schematic view to show the temperature variation of $d_{\rm 12}$ suggested from the present ARPES and band calculations.}
\end{figure}

As for a possible temperature variation of $d_{\rm 23}$, it is not excluded at this moment because the calculated band structure is insensitive to the value of $d_{\rm 23}$ [Fig. 4(c)]. It is noted that there exist some quantitative differences in the evolution of band structure between the experiment [Fig. 2(c)] and calculation [Fig. 4(b)], such as larger shift of the $k_{\rm F}$ point in the calculation. This may be related to the change in some parameters which was not considered in the calculation, like the lattice vibration involving in-plane ionic displacement. %%in the honeycomb lattice within the BL.

Now we discuss implications of the present result in relation to the previous experimental studies of Bi(111). From a quantitative comparison in the energy shift of the S2 band between the experiment and calculation in Figs. 2(c) and 4(b), it is inferred that $d_{\rm 12}$ is increased by $\sim$3\% (1.65/1.59 \AA) from 30 K to 300 K. On the other hand, a previous LEED $IV$ measurement on Bi(111) \cite{MonigPRB2005} suggested that the temperature dependence is weak for both $d_{\rm 12}$ and $d_{\rm 23}$. It was reported that, within the experimental uncertainty of LEED analysis, the $d_{\rm 12}$ and $d_{\rm 23}$ values are reduced by $\sim$1\% from 140 to 313 K. We think that the LEED and ARPES results are not necessarily incompatible with each other when taking into account the large error bar in the LEED $IV$ analysis ($\pm$3\% and $\pm$2\% at 313 K for $d_{\rm 12}$ and $d_{\rm 23}$, respectively) \cite{MonigPRB2005}. Also, the gradual expansion of $d_{\rm 12}$ with temperature observed in this study would be more reasonable in light of the positive thermal expansion coefficient of Bi crystal \cite{ErflingAP1939}. Thus, the present result suggests that temperature-dependent ARPES is highly sensitive to the change in structural parameters. We propose that the combination of temperature-dependent ARPES and DFT calculations works as an effective means to gain insight into the temperature variation of structural parameters at the surface. Since a similar structural relaxation of the out-of-plane lattice constant and a consequent temperature-dependent band shift may potentially take place in many other materials, application of this method to other systems is a next challenge in future.

\section{SUMMARY}
The present temperature-dependent ARPES study revealed an unexpectedly large variation of the band structure with temperature in Bi(111) ultrathin films. By comparing the observed temperature-induced band shift and DFT calculations for the Bi(111) slab, we suggest that the first interlayer spacing of Bi BL significantly expands on increasing temperature. The present study opens a new pathway toward studying the change in the structural parameters at the surface by combining temperature-dependent ARPES and DFT calculations.

\begin{acknowledgments}
This work was supported by Grant-in-Aid for Scientific Research on Innovative Areas ``Topological Materials Science'' (JSPS KAKENHI Grant Number JP15H05853 and No. JP18H04227), JST-CREST (No. JPMJCR18T1), Grant-in-Aid for Scientific Research (JSPS KAKENHI Grant Numbers JP17H01139, JP18H01160, JP19H01845, and JP18H01821), and KEK-PF (Proposal number 2018S2-001).
\end{acknowledgments}

\bibliographystyle{prsty}

\end{document}